%% file: HIN-23-007_temp.tex
\pdfoutput=1
\documentclass[11pt,twoside,a4paper,cmspaper,final,collab]{cms-tdr}
\def\svnVersion{5d6edbf}\def\svnDate{2025/01/21}\def\cmsCernNoTag{CERN-EP-2024-118}\def\cmsCernDate{\today}\def\cmsMessage{Published in Physics Letters B as \href{http://dx.doi.org/10.1016/j.physletb.2025.139279}{\doi{10.1016/j.physletb.2025.139279}.}}
\begin{document}\cmsNoteHeader{HIN-23-007}

\newlength\cmsTabSkip\setlength{\cmsTabSkip}{1ex}
\newlength\cmsFigWidth
\ifthenelse{\boolean{cms@external}}{\setlength\cmsFigWidth{0.99\columnwidth}}{\setlength\cmsFigWidth{0.6\textwidth}}
\ifthenelse{\boolean{cms@external}}{\newcommand{\cmsTop}{top\xspace}}{\newcommand{\cmsTop}{top\xspace}}
\ifthenelse{\boolean{cms@external}}{\newcommand{\cmsBottom}{bottom\xspace}}{\newcommand{\cmsBottom}{bottom\xspace}}
\ifthenelse{\boolean{cms@external}}{\newcommand{\cmsRight}{lower\xspace}}{\newcommand{\cmsRight}{right\xspace}}
\ifthenelse{\boolean{cms@external}}{\providecommand{\cmsTable}[1]{\resizebox{\columnwidth}{!}{#1}}}{\providecommand{\cmsTable}[1]{#1}}
\ifthenelse{\boolean{cms@external}}
{\providecommand{\suppMaterial}{the supplemental material [URL will be inserted by publisher]}}
{\providecommand{\suppMaterial}{Appendix~\ref{app:suppMat}}}

\newcommand {\nch}      {\ensuremath{N_{\text{ch}}}\xspace}
\newcommand {\dnchdeta} {\ensuremath{\rd N_{\text{ch}}/\rd \eta}\xspace}

\newcommand {\ampt}     {\textsc{ampt}\xspace}
\newcommand {\epos}     {\textsc{epos}\xspace}
\newcommand {\eposlhc}  {\textsc{epos lhc}\xspace}
\newcommand {\starlight}{\textsc{STARlight}\xspace}
\newcommand {\dpmjet}   {\textsc{dpmjet-III}\xspace}

\newcommand {\dr}       {\ensuremath{\Delta r}\xspace}
\newcommand {\vz}       {\ensuremath{v_{z}}\xspace}
\newcommand {\npart}    {\ensuremath{N_{\text{part}}}\xspace}
\newcommand {\avenpart} {\ensuremath{\langle N_{\text{part}} \rangle}\xspace}

\cmsNoteHeader{HIN-23-007}

\title{Pseudorapidity distributions of charged hadrons in lead-lead collisions at \texorpdfstring{$\sqrtsNN = 5.36\TeV$}{sqrt(s_NN) = 5.36 TeV}}

\date{\today}

\abstract{
The pseudorapidity ($\eta$) distributions of charged hadrons are measured using data collected at the highest ever nucleon-nucleon center-of-mass energy of $\sqrtsNN = 5.36\TeV$ for collisions of lead-lead ions.  The data were recorded by the CMS experiment at the LHC in 2022 and correspond to an integrated luminosity of $0.30\pm0.03\mubinv$.  Using the CMS silicon pixel detector, the yields of primary charged hadrons produced in the range $\abs{\eta}<2.6$ are reported.  The evolution of the midrapidity particle density as a function of collision centrality is also reported.  In the 5\% most central collisions, the charged-hadron $\eta$ density in the range $\abs{\eta} < 0.5$ is found to be $2032 \pm 91\syst$, with negligible statistical uncertainty.  This result is consistent with an extrapolation from nucleus-nucleus collision data at lower center-of-mass energies.  Comparisons are made to various Monte Carlo event generators and to previous measurements of lead-lead and xenon-xenon collisions at similar collision energies.  These new data detail the dependence of particle production on the collision energy, initial collision geometry, and the size of the colliding nuclei.
}

\hypersetup{
pdfauthor={CMS Collaboration},
pdftitle={Pseudorapidity distributions of charged hadrons in lead-lead collisions at 5.36 TeV},
pdfsubject={CMS},
pdfkeywords={CMS, hadrons, multiplicity, spectra}}

\maketitle
\section{Introduction}
A hot medium of deconfined quarks and gluons, known as the quark-gluon plasma (QGP), is created in high-energy heavy ion collisions~\cite{Busza:2018rrf}.  These collisions provide a unique avenue for the experimental study of matter whose properties are dictated by quantum chromodynamics (QCD).  

One of most interesting aspects of the QGP is its behavior as an ``almost perfect" liquid, \ie, having a very low shear viscosity to entropy density ratio, and the ability of hydrodynamic models to describe the evolution of the system~\cite{Heinz:2013th}.  Charged-hadron pseudorapidity ($\eta$) distributions and overall multiplicities (\nch) can be used to constrain the initial conditions and subsequent evolution of the medium in these models~\cite{Romatschke:2017ejr}.  Furthermore, information regarding nuclear shadowing and gluon saturation effects~\cite{Albacete:2014fwa} can be extracted by studying the dependence of the \nch on the initial geometry and center-of-mass energy of the collision system.  Other interesting questions regarding the relative contributions to particle production of hard and soft scattering processes~\cite{Kharzeev:2000ph} and the modeling of these various processes~\cite{dEnterria:2011twh} can be studied using these observables.  Therefore, the measurements of the \nch and its $\eta$ dependence are important observables for understanding the formation and properties of the QGP in heavy ion collisions. 

In 2022, the CMS experiment collected a sample of lead-lead (PbPb) collision events at the highest ever nucleon-nucleon center-of-mass energy of $\sqrtsNN=5.36\TeV$, corresponding to an integrated luminosity of $0.30\pm0.03\mubinv$.  The collection of these data marked the beginning of the Run 3 era (2022--2025) at the CERN LHC for heavy ion collisions and provides an opportunity to examine the center-of-mass energy dependence of charged-particle production. Previous measurements of copper-copper (CuCu)~\cite{PHOBOS:2007hfc} and gold-gold (AuAu)~\cite{PHOBOS:2005zhy,BRAHMS:2001llo} collisions at lower energies at RHIC, as well as xenon-xenon (XeXe) collisions at 5.44 TeV~\cite{CMS:2019gzk,Acharya:2018hhy} and PbPb collisions at 2.76 TeV~\cite{ALICE:2010mlf,CMS:2011aqh} and 5.02 TeV~\cite{ALICE:2015juo} at the LHC, have indicated an approximate power-law scaling of charged-particle production as a function of the center-of-mass energy. This scaling law, as well as models that have been tuned at lower energies~\cite{Deng:2010xg,Pierog:2013ria,Lokhtin:2005px}, can be tested using the new higher-energy data to determine how far the extrapolations from these empirical formulae and models can extend to higher center-of-mass energy systems. 

In this Letter, measurements of the pseudorapidity density, \dnchdeta, of primary
charged hadrons having $\abs{\eta} < 2.6$ are reported for PbPb collisions at $\sqrtsNN=5.36\TeV$. Primary charged hadrons
are defined as prompt charged hadrons and decay products of all particles with
proper decay length $c\tau < 1\unit{cm}$, where $c$ is the speed of light in
vacuum and $\tau$ is the proper lifetime of the particle.  The same definitions were used in previous analyses of
proton-proton ($\Pp\Pp$) collisions at 0.9--13\TeV~\cite{CMS:2010wcx,CMS:2010tjh,Khachatryan:2010nk,Chatrchyan:2014qka,CMS:2015zrm},
$\Pp$Pb collisions at 5.02 and 8.16\TeV~\cite{CMS:2017shj}, and PbPb
and XeXe collisions at 2.76 and 5.44\TeV~\cite{CMS:2011aqh,CMS:2019gzk}, respectively. Contributions from
prompt leptons, decay products of longer-lived particles, and secondary
interactions are excluded.  The measurements are reported as functions of charged hadron $\eta$ and the impact parameter between the two incoming ions in the collision.  The data are also compared to theoretical predictions from modern event generators.  Tabulated results are provided in the HEPData record for this analysis~\cite{hepdata}.

\section{The CMS detector}

The central feature of the CMS apparatus is a superconducting solenoid of
6\unit{m} internal diameter, providing a magnetic field of 3.8\unit{T}. Within
the solenoid volume are a silicon pixel and strip tracker covering the range
$\abs{\eta} < 2.5$, a lead tungstate crystal electromagnetic calorimeter, and a
brass and scintillator hadron calorimeter, each composed of a barrel and two
endcap sections. Forward calorimeters (HF), made of steel and quartz-fibers and
located on either side of the interaction point, extend the pseudorapidity
coverage provided by the barrel and endcap detectors to $\abs{\eta} < 5.2$.
Muons are detected in gas-ionization chambers embedded in the steel flux-return
yoke outside the solenoid. The beam pickup timing for experiments (BPTX)
devices are located around the beam pipe at a distance of 175\unit{m} from the
interaction point on either side and provide precise information on the timing
of the incoming beams. A more detailed description of the CMS detector,
together with a definition of the coordinate system used and the relevant
kinematic variables, can be found in Ref.~\cite{Chatrchyan:2008aa}.

Charged hadrons are reconstructed using the silicon pixel detector installed
during the Phase-1 upgrade~\cite{CMS:2012sda}, which consists of four concentric
cylindrical shells (layers) in the barrel region (BPIX) and three disks on each
side of the interaction point in the forward region (FPIX). The BPIX and FPIX
consist of a total of 1184 and 672 modules, respectively, and provide excellent
position resolution with their $100{\times}150\mum$ pixels. In this Letter, the
layers of the BPIX are denoted in increasing order of their radial distance
from the beam axis, \ie, the layer closest to the beam axis is referred to as
layer 1, the next closest layer is referred to as layer 2, and so on, while the
disks of the FPIX are referred to in increasing order of their longitudinal
distance from the nominal interaction point.

\section{Event selection}
\label{sec:eventselection}

Online, minimum-bias events were selected using a trigger that required a coincidence of signals from both BPTX detectors and at least one energy deposit above 3\GeV in each HF calorimeter.  Offline, events were required to have at least two energy deposits above 4\GeV in each of the HF calorimeters and at least one reconstructed primary vertex using the tracklet-based vertexing algorithm described in Ref.~\cite{CMS:2011aqh}. The tracklet reconstruction procedure is described in Section~\ref{sec:analysis}. A total of 50,000 events that pass these selections were used for the analysis.  
By analyzing data collected during the crossing of noncolliding ion bunches, the event selection requirements were found to successfully reject all backgrounds not arising from lead ion interactions.
Thus, beam backgrounds, beam-halo effects, and cosmic ray sources have a negligible impact on this analysis.

Although this analysis is predominantly focused on hadronic collisions, electromagnetic (EM) interactions between the lead ions must be accounted for.  The EM contribution is studied with simulated events generated by \starlight 2.2~\cite{Klein:2016yzr} interfaced with \dpmjet 3.0-5~\cite{Roesler:2000he}. These simulations estimate the contamination rate, which is included in the event selection efficiency uncertainty. Residual EM contributions are estimated from \starlight and cross-checked with data-driven methods that involve comparison between data and simulation with events that have small activities in the HF calorimeters. Single-diffractive events may also pass the event selections. Their contribution is subtracted using simulated events generated by \eposlhc v3400~\cite{Werner:2005jf,Pierog:2013ria}, in bins of tracklet multiplicity. The total hadronic event selection efficiency is calculated by comparing the total transverse energy distribution in the HF calorimeter with a template extracted from simulated \eposlhc events~\cite{CMS:2011aqh}.
The EM contamination rate is also factored into the total event selection efficiency uncertainty.

Because heavy ions are extended objects, their collisions can be characterized using the concept of ``centrality," which is related to the collision impact parameter.  This analysis estimates collision centrality by summing the transverse energy in the HF calorimeters, a procedure typically used in CMS analyses~\cite{Chatrchyan:2011sx,CMS:2011aqh}.  After correcting for the total event selection efficiency, the transverse energy distribution is partitioned into equal sections, which are labeled with a centrality percentage.  This percentage corresponds to a fraction of the total hadronic nuclear interaction cross section~\cite{CMS:2011aqh}.  By convention, the most central collisions, \ie, the ones having the smallest impact parameter, are labeled with small percentiles.  This analysis is restricted to collisions in the 0--80\% range, where the event selection efficiency is 100\% and EM contamination effects are small ($<$1\% contribution).
The collision centrality is correlated with the number of participating nucleons \npart in the event.  A Glauber model calculation~\cite{Miller:2007ri,Loizides:2017ack} is used to determine the average \npart that a given centrality range corresponds to.  These values are available in tabulated form in \suppMaterial.
For the purpose of this calculation, the nucleon-nucleon inelastic cross section is taken to be $68.0 \pm 1.2\unit{mb}$~\cite{Loizides:2017ack}. The nuclear radius and skin depth~\cite{Tarbert:2013jze} of the lead nucleus are set to $6.69 \pm 0.03$ and $0.56 \pm 0.03\unit{fm}$~\cite{Loizides:2014vua}, respectively.

\section{Analysis}
\label{sec:analysis}
The \dnchdeta measurement uses pairs of pixel clusters from two different layers (disks) of the silicon pixel detector. These clusters are formed from adjacent pixels with a charge above the readout threshold. These pairs, known as tracklets, have clusters with relatively small differences in $\eta$ and azimuthal angle $\phi$ (in radians) with respect to the primary vertex when they originate from a single charged particle.  The correlations in $\eta$ and $\phi$ can be used to select tracklets that correspond to primary charged hadrons.  Reference~\cite{CMS:2011aqh} contains more information on the tracklet and vertex reconstruction algorithms.

This analysis uses nine different types of tracklets, which are distinguished by the layers in the pixel detector that are used to form the cluster pair.  Six types result from using clusters in all possible combinations of the four barrel detector layers, and the remaining three come from combining pairs of the three forward disks. The combinations between barrel layers and forward disks are not used as they are sensitive to looper background from low transverse momentum ($\pt$) charged particles, which can have looping trajectories in the strong magnetic field and therefore leave multiple clusters in any given pixel layer.  The \dnchdeta distributions from all nine tracklet types, after applying the corrections discussed below, are averaged and symmetrized around $\eta = 0$ to yield the final \dnchdeta result. \dnchdeta distributions coming from each individual tracklet type have different sensitivities to the charged-hadron \pt spectrum and can therefore be used as a systematic check.  

A small fraction of detector modules were found to exhibit a significantly different response in the data when compared to Monte-Carlo (MC) simulations.  These detector modules are identified and excluded from subsequent analysis procedures. 

From the two clusters in a tracklet pair, a measure of angular distance 
\begin{equation}
\dr = \sqrt{{\left(\eta_{\mathrm{i}} - \eta_{\mathrm{j}}\right)^{2} + \left(\phi_{\mathrm{i}} - \phi_{\mathrm{j}}\right)^{2}}}
\end{equation}
is defined.  Here, $\eta_{\mathrm{i}(\mathrm{j})}$ is the pseudorapidity of the cluster in the
$\mathrm{i}(\mathrm{j})$th layer or disk with respect to the primary vertex position, and $\phi_{\mathrm{i}(\mathrm{j})}$ is defined similarly for the
azimuthal angle. The fraction of tracklets resulting from combinatorial background, formed by clusters originating from different charged particles, is estimated in the MC and found to dominate at $\dr > 0.5$ in the simulation. To increase the purity of the tracklet sample and to reduce the sensitivity to the modeling of combinatorial background effects, only tracklets having $\dr < 0.5$ are used in the analysis. This selection also suppresses the looper background related to low-$\pt$ charged particles. After the tracklet selection, the background rate (including misreconstructed tracklets and the particles that are not primary charged hadrons), geometric acceptance, reconstruction efficiency, and event selection efficiency of the detector are taken into account.  These corrections include an extrapolation to a tracklet \pt of zero to remove the effect of the minimum reconstructed charged-hadron \pt. The minimum charged-hadron \pt that can be reconstructed with this method is 40\MeV, corresponding to the tracklets formed using the inner two barrel pixel layers.
The correction factors are calculated using samples of the previously discussed MC generators that have been interfaced with \GEANTfour~\cite{Agostinelli:2002hh} to emulate the detector response. The typical magnitude of the corrections ranges from 30\% to 130\%. The tracklet efficiency shows minimal centrality dependence, but contamination rises in more central collisions due to acceptance effects and extrapolation to $\pt = 0$.

The \dr spectrum calculated with the \eposlhc sample is found to most closely match the data in the region where the combinatorial background is significant (\ie, $0.4<\dr <0.5$), so this sample is used for calculating the corrections for the nominal result in this analysis.  The simulated samples from other event generators are used to study systematic uncertainties.  The correction factors are calculated as functions of $\eta$, primary vertex position along the beam axis \vz, and tracklet multiplicity.

\section{Systematic uncertainties}

The uncertainties resulting from various systematic effects affecting the
measurement are evaluated:

\begin{itemize}
\item \textit{Different probability of pixel cluster splitting in data and simulation.}
Pixel cluster splitting refers to the condition where the charge deposit from a single charged
particle is reconstructed as two pixel clusters in close proximity. The
difference in the relative fraction of split clusters between data and
simulation can be estimated by artificially splitting the pixel clusters in
simulation and comparing the resulting modified \dr distribution of cluster
pairs in simulation to that in data. This difference in this relative fraction is found to be at most 2.0\%, which results in a variation of up to 2.0\% in the \dnchdeta results.
\item \textit{\vz consistency.} The potential mismodeling of the alignment of the pixel detector in the simulations affects
the correction factors. This effect is studied by comparing data with different \vz values which correspond to variations in $\eta$ acceptance. The variation, which depends on centrality and $\eta$, ranges from 0.5 to 4.5\%.
\item \textit{Tracklet selection.} The tracklet selection criteria affect the minimum \pt and signal-to-background
ratio of reconstructed tracklets. 
The sensitivity of correction factors to selection criteria is examined by adjusting the nominal \dr criterion by $\pm$0.1. This adjustment effectively acts as a filter for the minimum \pt, as particles with lower \pt exhibit larger \dr values. Specifically, changing the \dr cut modifies the minimum \pt by approximately 50\MeV for particles detected by combining the 1st and 4th layers in the barrel pixel detector. 
This uncertainty is less than 0.5\%.
\item \textit{Deviation from averaged and symmetrized results.} In any given $\eta$ range, measurements can be made
using multiple tracklet combinations. The maximum deviation in the measurements, obtained using each tracklet combination from the final averaged and symmetrized result, varies as a function of centrality and $\eta$, ranging from 1.0 to 4.0\%, and these values are used for the uncertainty.
\item \textit{Model dependence of the corrections.} The model dependence of the correction factors is studied by using
alternative sets of correction factors derived from \HYDJET 1.9~\cite{Lokhtin:2005px} and \ampt 1.26t5~\cite{Lin:2004en}, which have
different descriptions of the particle production mechanisms. The predicted
particle \pt distributions and the lepton fraction can differ significantly among the event
generators, which affect the correction for leptons and the extrapolation of
the measured tracklet spectra to $\pt = 0$. The maximum deviation from the
nominal results varies as a function of centrality and $\eta$, ranging from 0.5 to 2.0\%.
\item \textit{Centrality calibration.} The determination of event centrality depends on the hadronic event selection
efficiency, as well as the amount of contamination from EM processes. Since the
inefficiency is limited to the most peripheral collisions (events with the largest centrality), the effect of the
uncertainty in the event selection efficiency is to shift the events into other
centrality intervals. Hence, to evaluate the uncertainty in the final results,
different sets of centrality calibrations, derived after varying the event
selection efficiency by its uncertainty, are used to categorize the data.
This uncertainty increases from 0.1\% in central collisions to 2.5\% in peripheral events.
\end{itemize}
A summary of the systematic uncertainties affecting the measurement of charged hadron multiplicities in events with different mean number of participating nucleons ($\avenpart$) is given in Table~\ref{tab:tab1}.
The individual contributions are
then summed in quadrature to give the total systematic uncertainty, assuming them to be uncorrelated.  At forward $\eta$, the uncertainty related to the primary vertex position is the leading uncertainty.  At midrapidity, the dominant uncertainty comes from the consistency between tracklet combinations, except for peripheral events where the centrality calibration uncertainty is larger. 

\begin{table}[htbp!]
  \centering
    \topcaption{Sources of systematic uncertainty affecting the measurement of
    charged-hadron multiplicities as a function of \avenpart in PbPb collisions at $\sqrtsNN
    = 5.36\TeV$.}
    \label{tab:tab1}
    \cmsTable{
    \begin{tabular}{lc}
      Source                                          & [\%]          \\
    \hline
      Pixel cluster splitting                         & 1.4--2.0      \\
      Consistency between primary vertex position       & 0.5--4.5      \\
      Tracklet selection                              & ${<}0.5$           \\
      Consistency between tracklet combinations       & 1.0--4.0      \\
      Model dependence                                & 0.5--2.0      \\
      Centrality calibration                          & 0.1--2.5     \\[\cmsTabSkip]
     Total systematic uncertainty                     & 2.5--6.4 \\ 
    \end{tabular}
    }
\end{table}

\section{Results}

\begin{figure}[hbt!]
  \centering
    \includegraphics[width=0.45\textwidth]{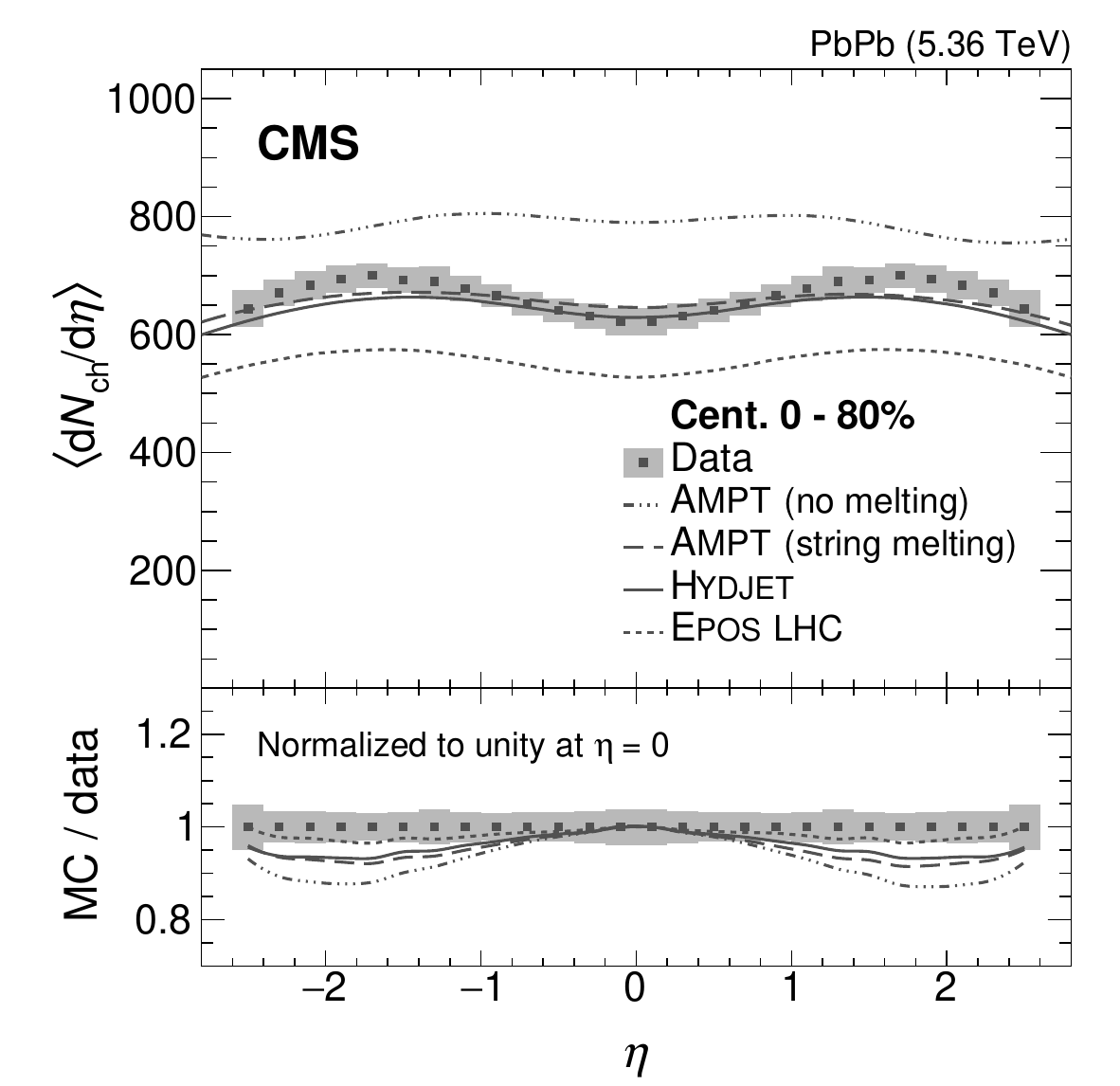}\\
    \includegraphics[width=0.45\textwidth]{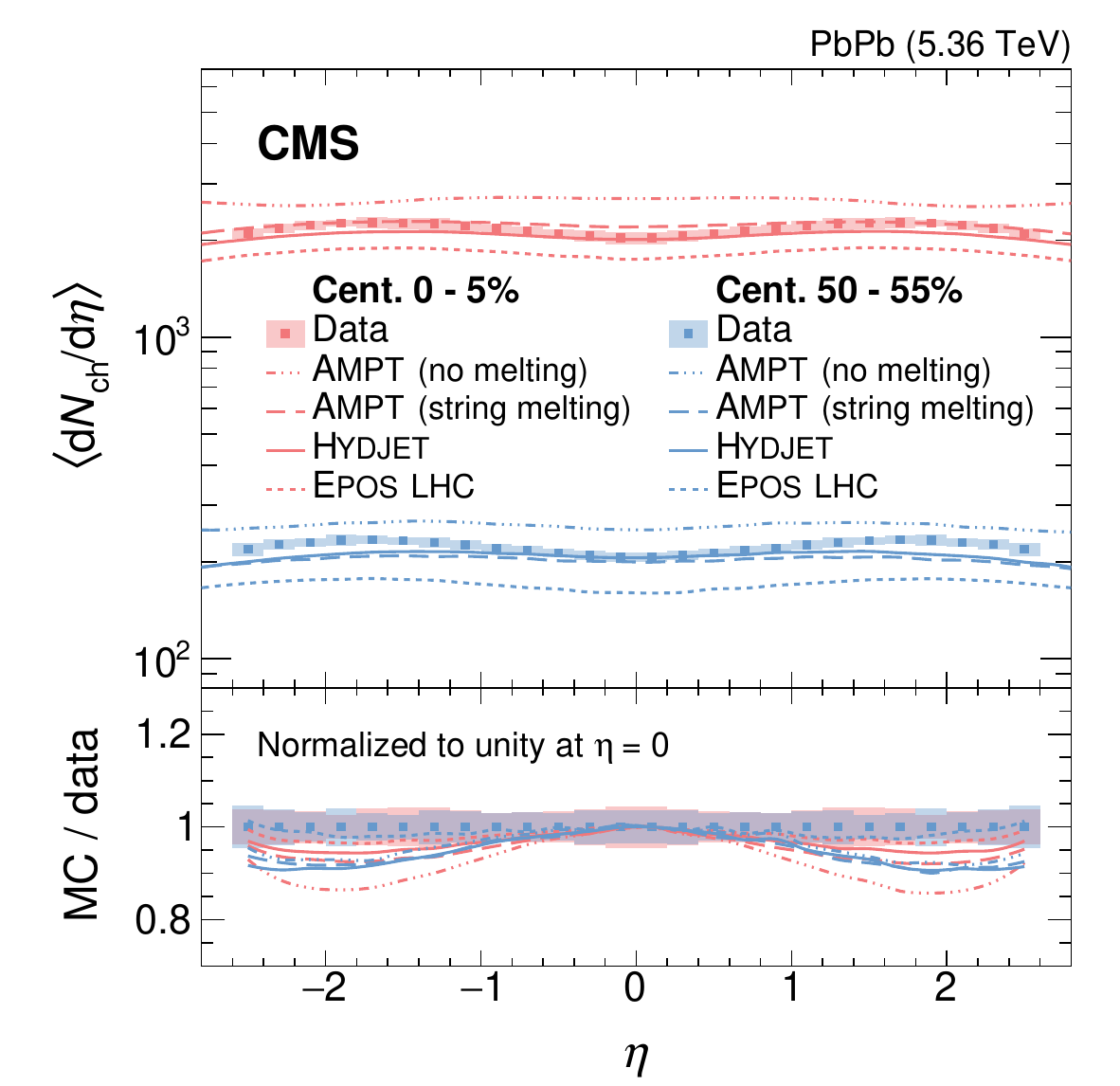}
    \caption{The \dnchdeta distributions in PbPb
    collisions at $\sqrtsNN = 5.36\TeV$ for events in the
    0--80\% centrality class (\cmsTop) and in the 0--5 and 50--55\% centrality classes (\cmsBottom).  The results have been averaged and symmetrized around $\eta=0$. Predictions
    from the \HYDJET 1.9~\cite{Lokhtin:2005px}, \ampt 1.26t5~\cite{Lin:2004en}, and \eposlhc v3400 event
    generators are also displayed.
    The ratios of the \dnchdeta distributions of simulation and data, normalized to unity at midrapidity, are shown in the bottom panels.
    The bands denote the total systematic uncertainties and the statistical uncertainties are negligible.
    In the ratio panels, the uncertainty band displayed represents the relative uncertainty of the data.}
    \label{fig:fig1}
\end{figure}

Figure~\ref{fig:fig1} (top) shows the charged-hadron $\eta$ distribution for the 0--80\% centrality class, while Fig.~\ref{fig:fig1} (bottom) presents this distribution for both the 0--5\% centrality class, indicating head-on collisions, and the 50--55\% class, which represents collisions with a larger impact parameter.

The results are compared to predictions from the \eposlhc
v3400, \HYDJET 1.9,
and \ampt 1.26t5 event generators. The \epos generator is
based on Gribov--Regge theory~\cite{Gribov:1968fc,Drescher:2000ha} and includes
the effect of collective hadronization in hadron-hadron scatterings. The \HYDJET
generator treats a heavy ion collision as a superposition of a hydrodynamically
parametrized soft component and a hard component resulting from multi-parton
fragmentation. The \ampt generator combines the \HIJING event
generator~\cite{Wang:1991hta} with Zhang's parton cascade
procedure~\cite{Zhang:1997ej} and the ART model~\cite{Li:1995pra} for the last
stage of parton hadronization.
For the 0--80\% event class and midrapidity region, the data match predictions from both \ampt with string melting and \HYDJET. The string melting mechanism in \ampt transforms all excited strings from heavy-ion collisions into partons and employs a spatial quark coalescence model for hadronization.

Similar conclusions hold for 0--5\% events when comparing to \HYDJET, but the agreement with \ampt with string melting seems to be better in the forward region than in the midrapidity region in this centrality interval.
The prediction of \ampt without string melting systematically overpredicts the data for all the centrality classes in the midrapidity region, while the predictions of \eposlhc significantly undershoot the data for all $\eta$ values and all of the centrality selections studied.  The lower panels of Fig.~\ref{fig:fig1} show the ratios of the MC models to data with the normalization set such that the ratio is unity at $\eta = 0$.  It can be seen that, after normalization, differences are accounted for, \eposlhc does the best job of all the models examined of predicting the shape of the $\eta$ distribution.  Similar conclusions were found in studies of XeXe collisions at $\sqrtsNN = 5.44\TeV$ by the CMS~\cite{CMS:2019gzk} and ALICE~\cite{Acharya:2018hhy} Collaborations.  Nonetheless, none of the MC generators examined here are able to describe both the shape and magnitude of the charged-hadron distribution across the full range of $\eta$ probed.

\begin{figure}[hbt!]
  \centering
    \includegraphics[width=0.45\textwidth]{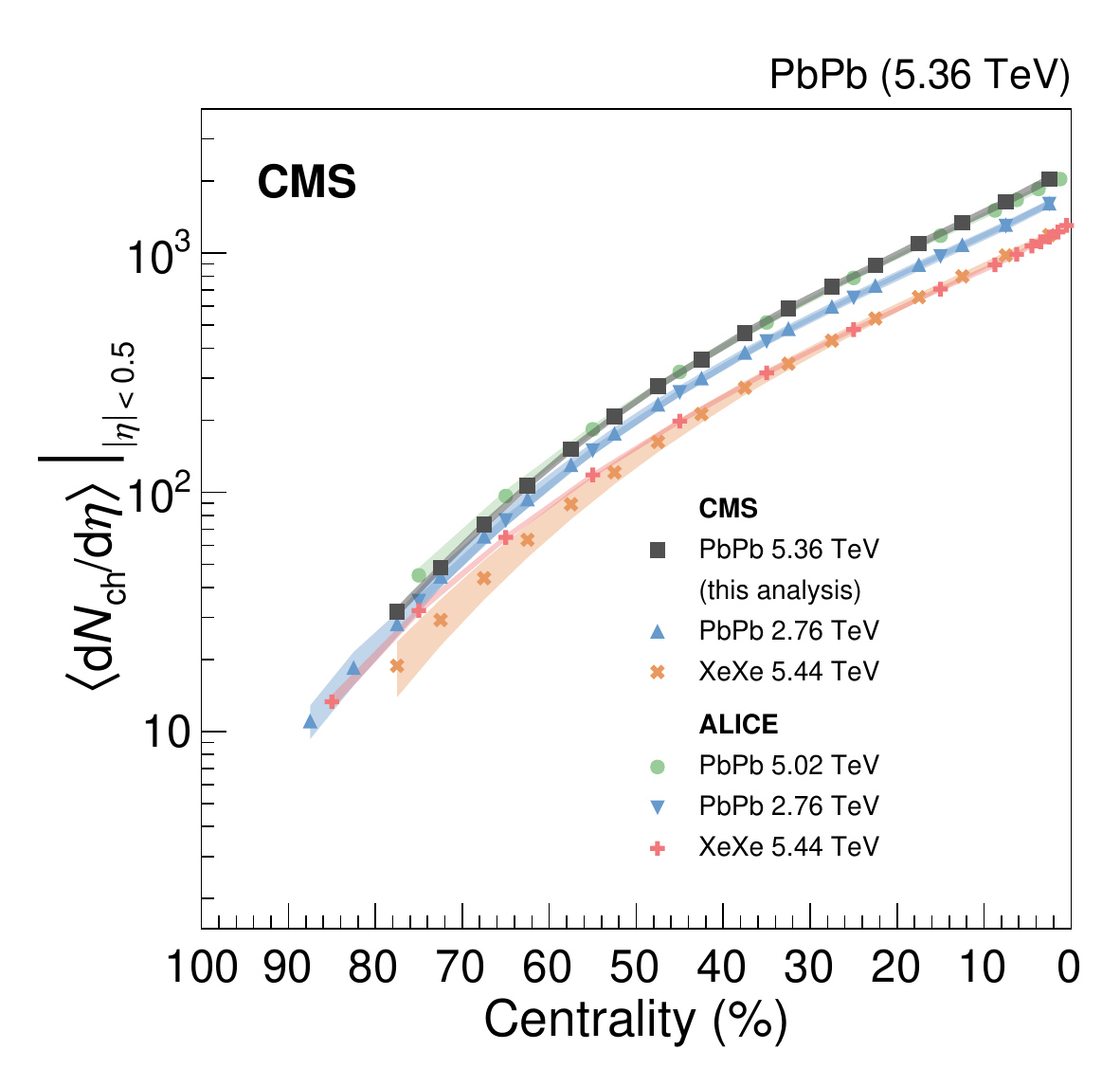}\\
    \includegraphics[width=0.45\textwidth]{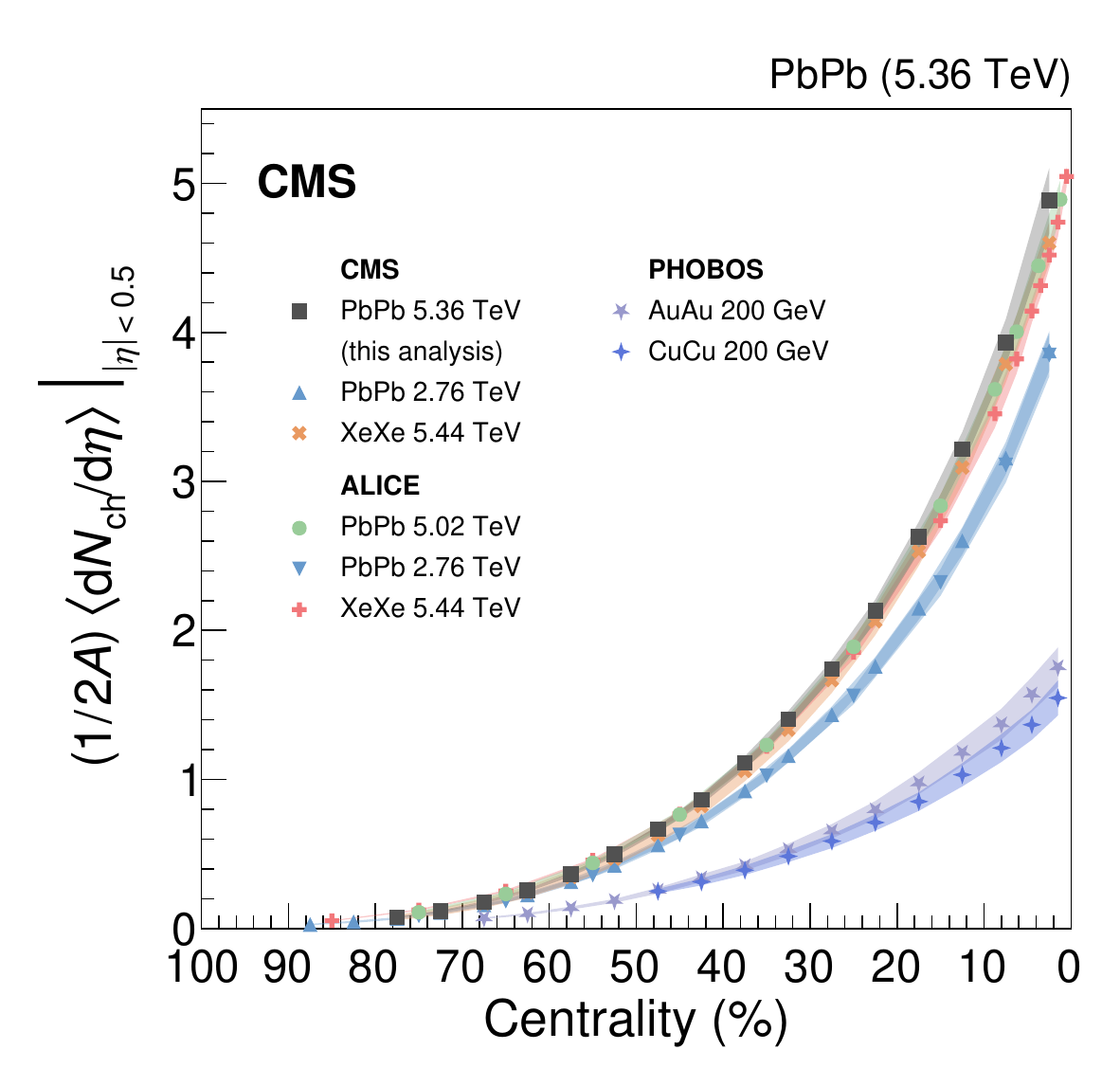}
    \caption{Charged-hadron \dnchdeta in PbPb collisions at $\sqrtsNN =
    5.36\TeV$ at midrapidity as a function of event centrality, shown as is
    (\cmsTop) and normalized by $2A$ (\cmsBottom), where $A$ is the atomic
    number of the nuclei. The results are compared to measurements in PbPb and
    XeXe collisions by the CMS~\cite{CMS:2011aqh,CMS:2019gzk} and
    ALICE~\cite{ALICE:2010khr,ALICE:2015juo,Acharya:2018hhy} Collaborations, and
    to measurements in CuCu and AuAu collisions by the PHOBOS
    Collaboration~\cite{PHOBOS:2010eyu}. The bands around the data points denote
    the total systematic uncertainties, while the statistical uncertainties are
    negligible.}
    \label{fig:fig2}
\end{figure}

The charged-hadron \dnchdeta at
midrapidity as a function of centrality is shown in Fig.~\ref{fig:fig2} (top). The value of \dnchdeta is found to be $2032 \pm 91\syst$ ($207 \pm 7\syst$) at $\abs{\eta}<0.5$ for events in the 0--5 (50--55\%) centrality class.
Previous measurements of PbPb collisions at $\sqrtsNN = 2.76$ and 5.02\TeV by the CMS~\cite{CMS:2011aqh} and
ALICE~\cite{ALICE:2010khr,ALICE:2015juo} Collaborations, as well as previously discussed measurements of XeXe collisions at $5.44\TeV$~\cite{CMS:2019gzk,Acharya:2018hhy}, are also shown.
In general, at the same centrality \dnchdeta is larger for the collision systems with heavier ions. The PbPb data show a higher charged hadron \dnchdeta at higher collision energy for central events, as expected.  The new 5.36\TeV CMS data are consistent with the ALICE data from 5.02\TeV collisions within uncertainties.

Previous analyses have noted that \dnchdeta approximately scales with $2A$ for collisions of different ion species at the same collision energy~\cite{CMS:2019gzk}, where $A$ represents the atomic number of the colliding nuclei. Figure~\ref{fig:fig2} (bottom) displays the normalized particle density $(\dnchdeta)/2A$ as a function of collision centrality for the same data sets shown in Fig.~\ref{fig:fig2} (top).  Although the 5.36\TeV data are closer in collision energy to the 5.44\TeV XeXe data than the previous 5.02\TeV PbPb data, they tend to have slightly higher normalized \dnchdeta than 5.02\TeV PbPb, indicating that this scaling law is only approximate.

\begin{figure}[hbt!]
  \centering
    \includegraphics[width=0.45\textwidth]{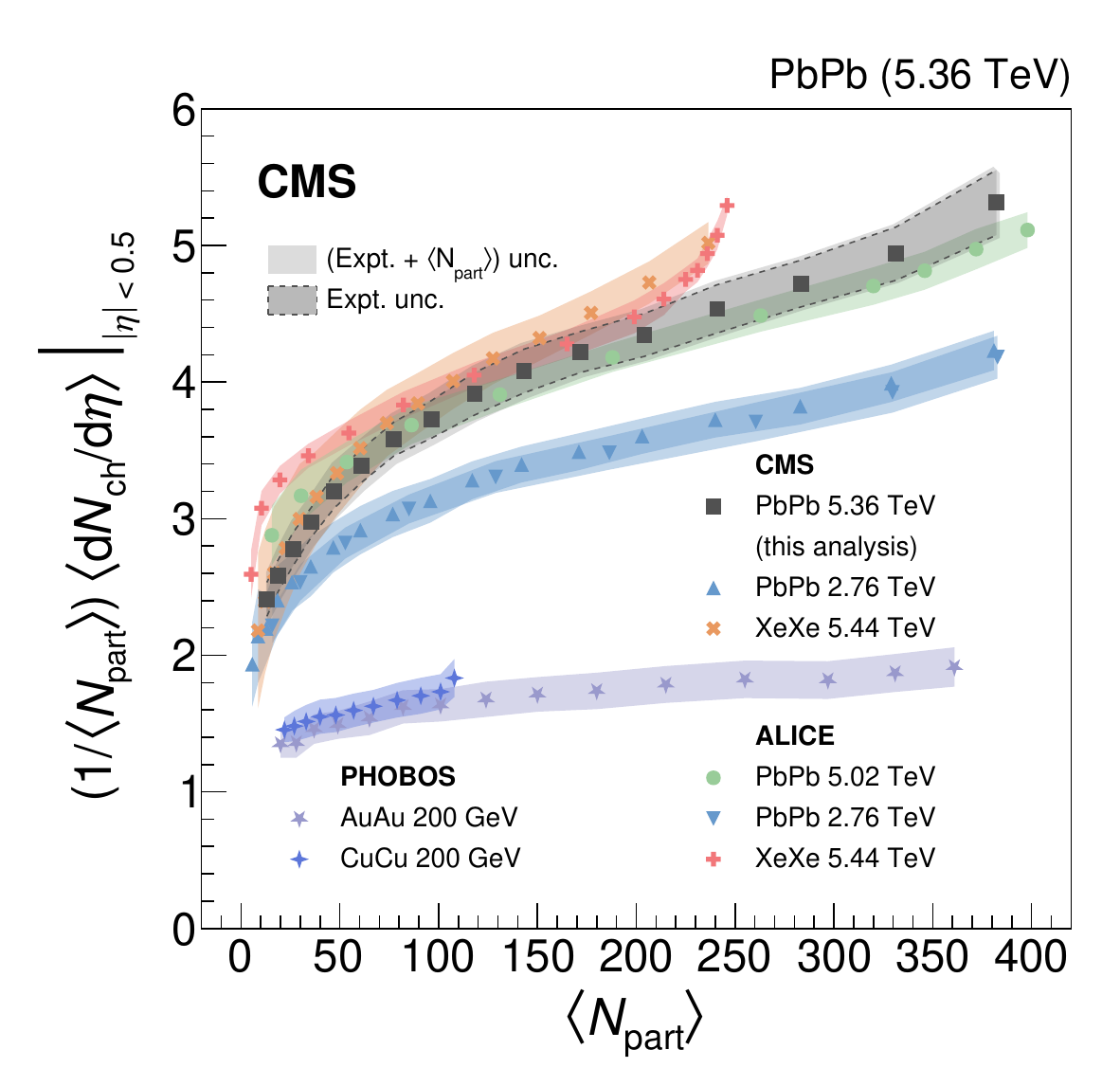}\\
    \includegraphics[width=0.45\textwidth]{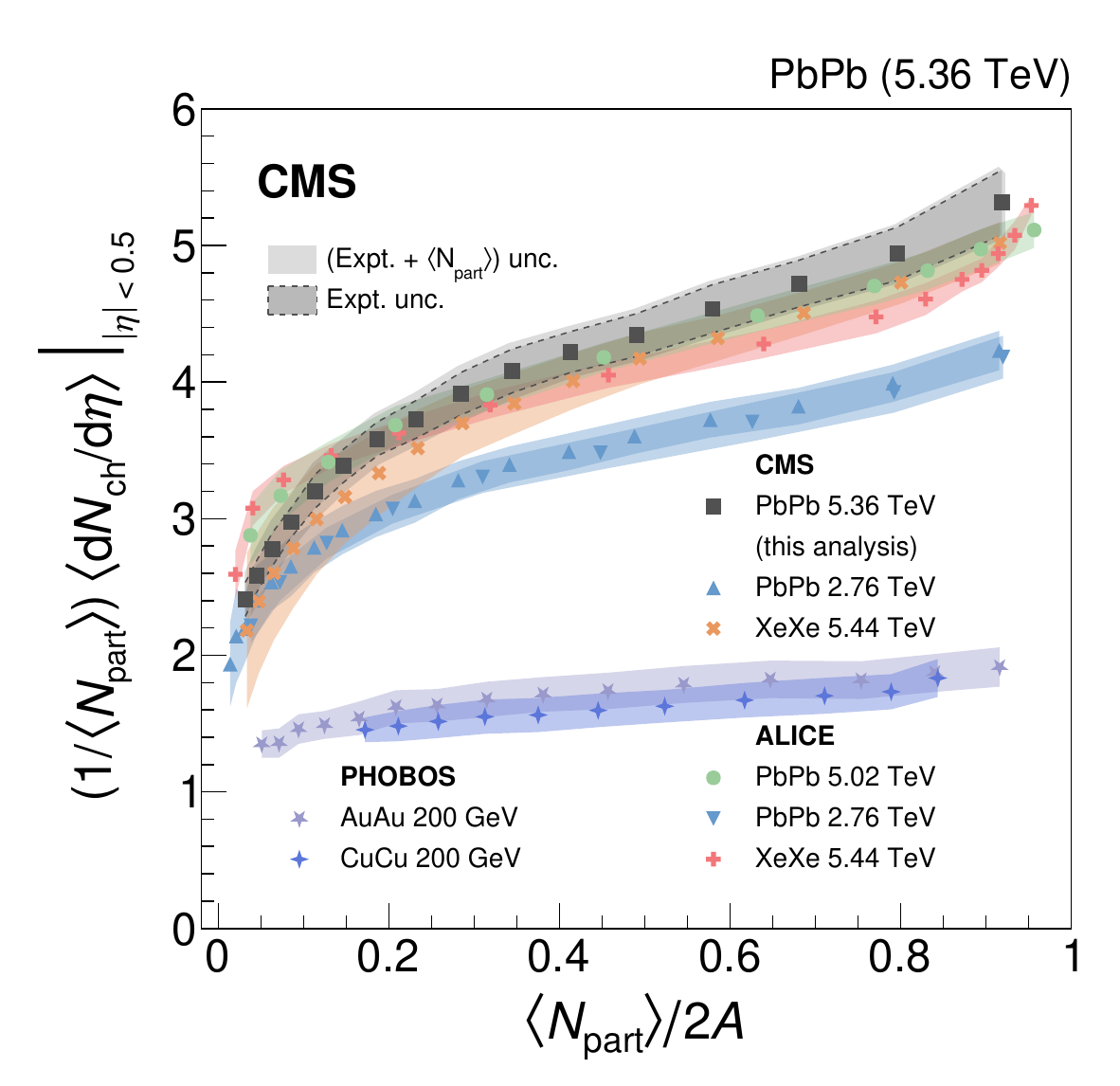}
    \caption{Average \dnchdeta at midrapidity normalized by \avenpart, shown as
    a function of \avenpart (\cmsTop) and $\avenpart/2A$ (\cmsBottom), where
    $A$ is the atomic number of the nuclei. The results are compared to
    measurements in PbPb and XeXe collisions by the CMS~\cite{CMS:2011aqh,CMS:2019gzk},
     ALICE~\cite{ALICE:2010khr,ALICE:2015juo,Acharya:2018hhy} Collaborations, and
    to measurements in CuCu and AuAu collisions by the PHOBOS
    Collaboration~\cite{PHOBOS:2010eyu}.
    The bands around the data points denote the systematic uncertainties, while
    the statistical uncertainties are negligible.}
    \label{fig:fig3}
\end{figure}

The value of $(\dnchdeta)/\avenpart$ is shown as a function of \avenpart in Fig.~\ref{fig:fig3} (\cmsTop)
to study the role of the number of participating nucleons in particle production.
The results are also compared to lower-energy measurements of CuCu and AuAu collisions at $\sqrtsNN = 200\GeV$, in addition to those from LHC experiments.
It is important to note that the same number of participants can correspond to different centrality classes across various collision systems. The measured multiplicity per participant increases with higher collision energy for events with a similar \avenpart. 
At similar collision energy, the shapes of the distributions are inconsistent between PbPb and XeXe collisions.  This is particularly apparent when examining the behavior around $\avenpart = 200$, corresponding to the most central XeXe collisions.
However, as shown in Fig.~\ref{fig:fig3} (\cmsBottom), the per-participant \nch for
different colliding nuclei is consistent within uncertainties when the fraction of nucleons participating
($\avenpart/2A$) and energy of the compared systems are the same.  These data clearly suggest that the initial collision geometry is a crucial factor when predicting particle production at LHC energies~\cite{PHOBOS:2007hfc}.

\begin{figure}[hbt!]
  \centering
     \includegraphics[width=\cmsFigWidth]{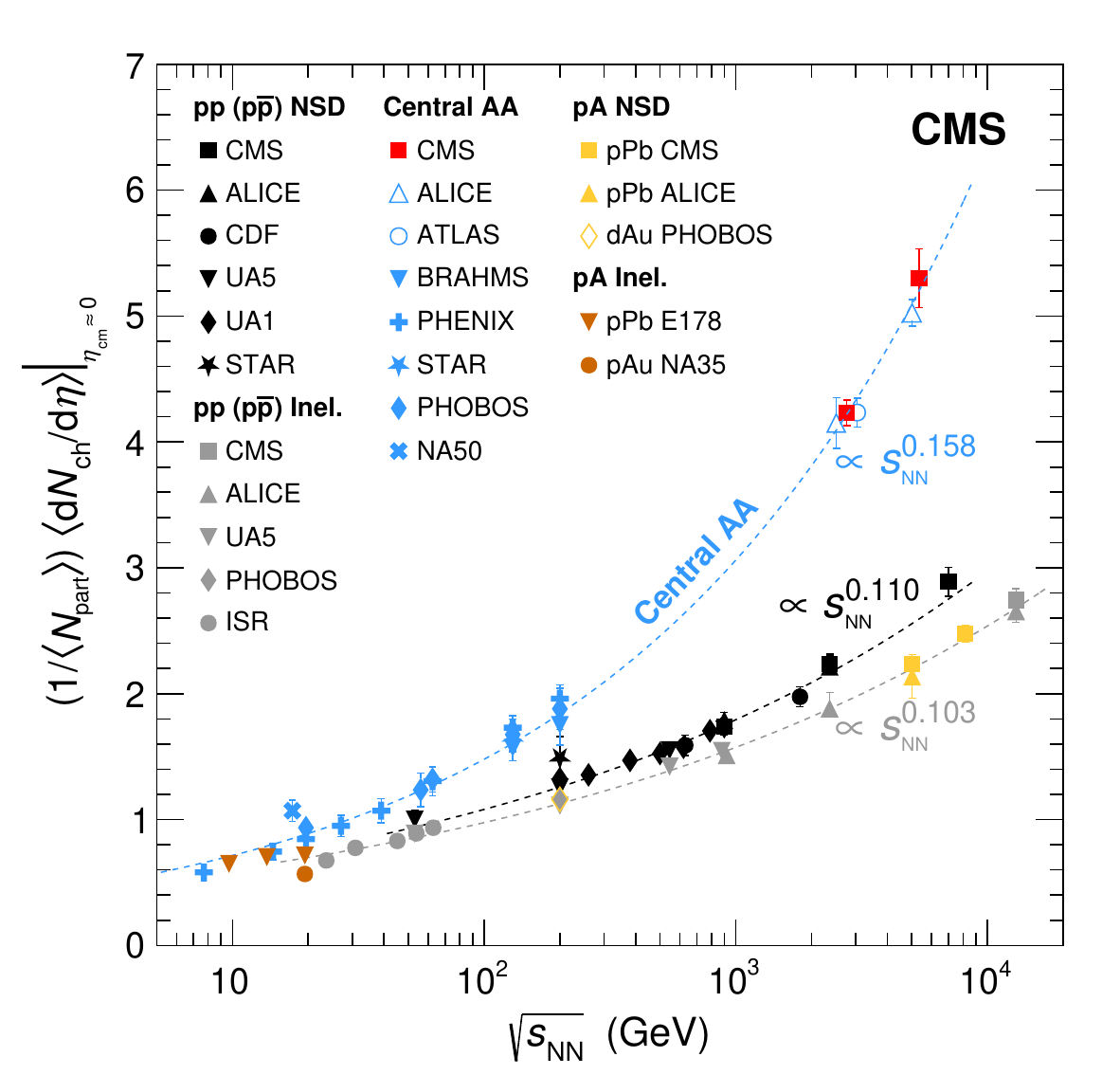}
    \caption{Comparison of average \dnchdeta at midrapidity, scaled by \avenpart in pPb~\cite{ALICE:2012xs,CMS:2017shj}, pAu~\cite{NA35:1997dow}, dAu~\cite{PHOBOS:2003fjw,PHENIX:2015tbb,PHENIX:2017nae} (pA) and central heavy ion collisions~\cite{CMS:2011aqh,PHOBOS:2010eyu,NA50:2002edr,STAR:2001eyo,BRAHMS:2001gci,BRAHMS:2001llo,PHENIX:2000owy,PHOBOS:2000wxz,PHOBOS:2001zjw,ALICE:2010khr,ATLAS:2011ag,PHENIX:2015tbb,STAR:2009sxc,STAR:2008med}, as well as non-single-diffractive (NSD)~\cite{CMS:2010wcx,CMS:2010tjh,STAR:2008med,UA1:1989bou,UA5:1982yte,CDF:1989nkn} and inelastic~\cite{CMS:2015zrm,PHOBOS:2010eyu,Aachen-CERN-Heidelberg-Munich:1977izz,ALICE:2010cin} pp collisions.
The data points for nucleus-nucleus (AA) collisions at $\sqrtsNN = 2.76\TeV$ have been shifted horizontally for visibility. The dashed
curves, reproduced from Ref.~\cite{CMS:2017shj}, are included to guide the eye, and correspond to a power law functional form.}
    \label{fig:fig4}
\end{figure}

Figure~\ref{fig:fig4} shows the dependence of \dnchdeta normalized by \avenpart on the collision energy for
various collision systems and event selections.  Power law fits to the previous measurements reproduced from Ref.~\cite{CMS:2017shj} are displayed using dashed lines to guide the eye.  The new CMS data, shown by the red point at $\sqrtsNN=5.36\TeV$, are consistent with this extrapolation from lower ion-ion collision energies; the central PbPb results at 5.36\TeV are about two times larger than non-single-diffractive (NSD) $\Pp\Pp$ collisions.  It has been previously noted in Ref.~\cite{CMS:2017shj} that NSD pPb collisions (shown by yellow markers) have a lower normalized charged-particle density compared to NSD $\Pp\Pp$ collisions (black markers) at the same energy, potentially because of the presence of non-QGP related effects.  Although such effects may also be present in ion-ion collisions to some extent, the different trends of ion-ion and proton-ion ($\Pp$A) data with respect to NSD $\Pp\Pp$ collisions highlight the complexity of the interplay between non-QGP and QGP-related effects when considering soft-particle production.  These data clearly indicate that ion-ion collisions are more efficient than smaller collision systems at converting the initial collision energy into final-state particle multiplicity, and they are not merely superpositions of $\Pp\Pp$ or $\Pp$A collisions.

\section{Summary}

The pseudorapidity ($\eta$) distributions of charged hadrons are measured in the range $\abs{\eta}<2.6$ for multiple centrality intervals using data collected in lead-lead (PbPb) collisions at the highest center-of-mass energy of $\sqrtsNN = 5.36\TeV$.
The dependence on $\eta$ and centrality are compared to the event generators \eposlhc v3400, \HYDJET 1.9, and \ampt 1.26t5; none of these event generators are able to fully describe the
measurements in terms of the magnitude, $\eta$ dependence, and
centrality dependence of the \dnchdeta distributions.
The charged hadron density at midrapidity rises with collision energy and centrality, aligning with extrapolations from lower energies. Additionally, the scaling of this observable with the size of the colliding nuclei is found to be only approximate.

In the 5\% most central collisions, the charged-hadron density \dnchdeta for the range $\abs{\eta} < 0.5$ is found to be $2032 \pm 91\syst$, with negligible statistical uncertainty.
The results at midrapidity are also compared to previous measurements in various collision systems, including PbPb collisions at $\sqrtsNN = 2.76$ and 5.02\TeV, xenon-xenon collisions at 5.44\TeV, as well as copper-copper and gold-gold collisions at $\sqrtsNN = 200\GeV$. Normalizing \dnchdeta by the number of participating nucleons shows that larger systems convert collision energy to particles more efficiently than smaller ones. These comparisons are new constraints on models and generators which describe multiparticle production in relativistic heavy ion collisions.

\begin{acknowledgments}
  We congratulate our colleagues in the CERN accelerator departments for the excellent performance of the LHC and thank the technical and administrative staffs at CERN and at other CMS institutes for their contributions to the success of the CMS effort. In addition, we gratefully acknowledge the computing centers and personnel of the Worldwide LHC Computing Grid and other centers for delivering so effectively the computing infrastructure essential to our analyses. Finally, we acknowledge the enduring support for the construction and operation of the LHC, the CMS detector, and the supporting computing infrastructure provided by the following funding agencies: SC (Armenia), BMBWF and FWF (Austria); FNRS and FWO (Belgium); CNPq, CAPES, FAPERJ, FAPERGS, and FAPESP (Brazil); MES and BNSF (Bulgaria); CERN; CAS, MoST, and NSFC (China); MINCIENCIAS (Colombia); MSES and CSF (Croatia); RIF (Cyprus); SENESCYT (Ecuador); ERC PRG, RVTT3 and MoER TK202 (Estonia); Academy of Finland, MEC, and HIP (Finland); CEA and CNRS/IN2P3 (France); SRNSF (Georgia); BMBF, DFG, and HGF (Germany); GSRI (Greece); NKFIH (Hungary); DAE and DST (India); IPM (Iran); SFI (Ireland); INFN (Italy); MSIP and NRF (Republic of Korea); MES (Latvia); LMTLT (Lithuania); MOE and UM (Malaysia); BUAP, CINVESTAV, CONACYT, LNS, SEP, and UASLP-FAI (Mexico); MOS (Montenegro); MBIE (New Zealand); PAEC (Pakistan); MES and NSC (Poland); FCT (Portugal); MESTD (Serbia); MCIN/AEI and PCTI (Spain); MOSTR (Sri Lanka); Swiss Funding Agencies (Switzerland); MST (Taipei); MHESI and NSTDA (Thailand); TUBITAK and TENMAK (Turkey); NASU (Ukraine); STFC (United Kingdom); DOE and NSF (USA).

  \hyphenation{Rachada-pisek} Individuals have received support from the Marie-Curie program and the European Research Council and Horizon 2020 Grant, contract Nos.\ 675440, 724704, 752730, 758316, 765710, 824093, 101115353, 101002207, and COST Action CA16108 (European Union); the Leventis Foundation; the Alfred P.\ Sloan Foundation; the Alexander von Humboldt Foundation; the Science Committee, project no. 22rl-037 (Armenia); the Belgian Federal Science Policy Office; the Fonds pour la Formation \`a la Recherche dans l'Industrie et dans l'Agriculture (FRIA-Belgium); the F.R.S.-FNRS and FWO (Belgium) under the ``Excellence of Science -- EOS" -- be.h project n.\ 30820817; the Beijing Municipal Science \& Technology Commission, No. Z191100007219010 and Fundamental Research Funds for the Central Universities (China); the Ministry of Education, Youth and Sports (MEYS) of the Czech Republic; the Shota Rustaveli National Science Foundation, grant FR-22-985 (Georgia); the Deutsche Forschungsgemeinschaft (DFG), among others, under Germany's Excellence Strategy -- EXC 2121 ``Quantum Universe" -- 390833306, and under project number 400140256 - GRK2497; the Hellenic Foundation for Research and Innovation (HFRI), Project Number 2288 (Greece); the Hungarian Academy of Sciences, the New National Excellence Program - \'UNKP, the NKFIH research grants K 131991, K 133046, K 138136, K 143460, K 143477, K 146913, K 146914, K 147048, 2020-2.2.1-ED-2021-00181, and TKP2021-NKTA-64 (Hungary); the Council of Science and Industrial Research, India; ICSC -- National Research Center for High Performance Computing, Big Data and Quantum Computing and FAIR -- Future Artificial Intelligence Research, funded by the NextGenerationEU program (Italy); the Latvian Council of Science; the Ministry of Education and Science, project no. 2022/WK/14, and the National Science Center, contracts Opus 2021/41/B/ST2/01369 and 2021/43/B/ST2/01552 (Poland); the Funda\c{c}\~ao para a Ci\^encia e a Tecnologia, grant CEECIND/01334/2018 (Portugal); the National Priorities Research Program by Qatar National Research Fund; MCIN/AEI/10.13039/501100011033, ERDF ``a way of making Europe", and the Programa Estatal de Fomento de la Investigaci{\'o}n Cient{\'i}fica y T{\'e}cnica de Excelencia Mar\'{\i}a de Maeztu, grant MDM-2017-0765 and Programa Severo Ochoa del Principado de Asturias (Spain); the Chulalongkorn Academic into Its 2nd Century Project Advancement Project, and the National Science, Research and Innovation Fund via the Program Management Unit for Human Resources \& Institutional Development, Research and Innovation, grant B39G670016 (Thailand); the Kavli Foundation; the Nvidia Corporation; the SuperMicro Corporation; the Welch Foundation, contract C-1845; and the Weston Havens Foundation (USA).
\end{acknowledgments}

\bibliography{auto_generated}

\ifthenelse{\boolean{cms@external}}{}{
\clearpage
\appendix
\numberwithin{table}{section}
\section{
\texorpdfstring{Supplemental material: Centrality intervals and corresponding \avenpart values}
{Supplemental material: Centrality intervals and corresponding mean Npart values}
\label{app:suppMat}}
\input{supplemental_material}
}

\cleardoublepage \section{The CMS Collaboration \label{app:collab}}\begin{sloppypar}\hyphenpenalty=5000\widowpenalty=500\clubpenalty=5000\input{HIN-23-007-public-authorlist.tex}\end{sloppypar}
\end{document}

%% file: supplemental_material.tex
\begin{table}[htbp!]
  \centering
  \topcaption{Centrality intervals and corresponding \avenpart
      values for 5.36\TeV PbPb collisions.  The uncertainties in the \npart values are determined by propagating the uncertainties in the parameters of the Glauber model.}
    \label{tab:tab2}
    \begin{tabular}{cc}
      Centrality interval [\%] & \avenpart \\ 
    \hline
    0--5 & $382.3 \pm 1.6$ \\
    5--10 & $331.3 \pm 1.3$ \\
    10--15 & $283.3 \pm 1.4$ \\
    15--20 & $241.0 \pm 1.4$ \\
    20--25 & $204.1 \pm 1.5$ \\
    25--30 & $171.7 \pm 1.5$ \\
    30--35 & $143.2 \pm 1.5$ \\
    35--40 & $118.2 \pm 1.4$ \\
    40--45 & $96.3 \pm 1.3$ \\
    45--50 & $77.4 \pm 1.3$ \\
    50--55 & $61.1 \pm 1.2$ \\
    55--60 & $47.2 \pm 1.2$ \\
    60--65 & $35.7 \pm 0.9$ \\
    65--70 & $26.3 \pm 0.8$ \\
    70--75 & $18.8 \pm 0.6$ \\
    75--80 & $13.1 \pm 0.4$ \\
    \end{tabular}
\end{table}

%% file: HIN-23-007-public-authorlist.tex
\cmsinstitute{Yerevan Physics Institute, Yerevan, Armenia}
{\tolerance=6000
A.~Hayrapetyan, A.~Tumasyan\cmsAuthorMark{1}\cmsorcid{0009-0000-0684-6742}
\par}
\cmsinstitute{Institut f\"{u}r Hochenergiephysik, Vienna, Austria}
{\tolerance=6000
W.~Adam\cmsorcid{0000-0001-9099-4341}, J.W.~Andrejkovic, T.~Bergauer\cmsorcid{0000-0002-5786-0293}, S.~Chatterjee\cmsorcid{0000-0003-2660-0349}, K.~Damanakis\cmsorcid{0000-0001-5389-2872}, M.~Dragicevic\cmsorcid{0000-0003-1967-6783}, P.S.~Hussain\cmsorcid{0000-0002-4825-5278}, M.~Jeitler\cmsAuthorMark{2}\cmsorcid{0000-0002-5141-9560}, N.~Krammer\cmsorcid{0000-0002-0548-0985}, A.~Li\cmsorcid{0000-0002-4547-116X}, D.~Liko\cmsorcid{0000-0002-3380-473X}, I.~Mikulec\cmsorcid{0000-0003-0385-2746}, J.~Schieck\cmsAuthorMark{2}\cmsorcid{0000-0002-1058-8093}, R.~Sch\"{o}fbeck\cmsorcid{0000-0002-2332-8784}, D.~Schwarz\cmsorcid{0000-0002-3821-7331}, M.~Sonawane\cmsorcid{0000-0003-0510-7010}, S.~Templ\cmsorcid{0000-0003-3137-5692}, W.~Waltenberger\cmsorcid{0000-0002-6215-7228}, C.-E.~Wulz\cmsAuthorMark{2}\cmsorcid{0000-0001-9226-5812}
% ...
\par}
\vskip\cmsinstskip
\dag:~Deceased\\
$^{1}$Also at Yerevan State University, Yerevan, Armenia\\
$^{2}$Also at TU Wien, Vienna, Austria\\
% ...
